\documentclass{PoS}

\title{
{\vspace{-20mm} \normalsize
\hfill{\small MS-TP-07-28 }}    \\[25mm] 
			       Strong coupling expansion for Yang-Mills theory at finite temperature}

\ShortTitle{Strong coupling expansion for Yang-Mills theory at finite temperature}

\author{\speaker{Jens Langelage}, Gernot M\"unster, Owe Philipsen \\
        Westf\"alische Wilhelms-Universit\"at M\"unster, Institut f\"ur Theoretische Physik, Wilhelm-Klemm-Str.\,9, 48149 M\"unster, Germany\\
        E-mail: jens.langelage@uni-muenster.de\\
	E-mail: munsteg@uni-muenster.de\\
	E-mail: ophil@uni-muenster.de}

\abstract{
Euclidean strong coupling expansion of the partition function is applied
to lattice Yang-Mills theory at finite temperature, i.e. for lattices
with a compactified temporal direction.
The expansions have a finite radius of convergence 
and thus are valid only for $\beta<\beta_c$, where $\beta_c$ denotes the nearest
singularity of the free energy on the real axis.
The accessible temperature range is thus the confined regime
up to the deconfinement transition. 
We have calculated the first few orders of these expansions of the free 
energy density as well as the screening masses for the 
gauge groups SU(2) and SU(3). 
The resulting free energy series can be summed up and corresponds 
to a glueball gas of the lowest mass glueballs
up to the calculated order.
Our result can be used to fix the lower integration constant for Monte
Carlo calculations of the thermodynamic pressure via the integral method,
and shows from first principles that in the confined phase
this constant is indeed exponentially small.
Similarly, our results also explain the weak temperature dependence of 
glueball screening masses below $T_c$, as observed in Monte Carlo simulations.
Possibilities and difficulties in extracting $\beta_c$ from the series
are discussed.}

\FullConference{The XXV International Symposium on Lattice Field Theory\\
		 July 30-4 August 2007\\
		 Regensburg, Germany}

\begin{document}

\section{Introduction}

Contrary to weak coupling expansions, strong coupling expansions are known to 
be convergent series with a finite
radius of convergence. In the early days of lattice gauge theory they were 
used to get analytical results for some physical quantities of interest, 
such as glueball masses or the energy density of lattice Yang-Mills theories. 
These calculations were done at zero temperature, i.e.~at infinite 
volume $N_s^3$ and temporal extent 
$N_t$ of the lattice.

Here we calculate such series expansions for the free energy density 
and screening masses with an infinite 
spatial volume and a compactified temporal extension $N_t$ of the lattice. 
In this way finite temperature effects are generated, giving us the opportunity to study the physical, 
temperature dependent free energy density in the confined phase. The physical deconfinement phase transition
then corresponds to a finite convergence radius of the series, which one may try to estimate
from the behaviour of the coefficients.

\section{Free energy density}

\subsection{Cluster expansion}

The partition function of the lattice Yang-Mills theory is given by a functional 
integration of the exponentiated Wilson action
over the corresponding SU(N) group space,
\begin{eqnarray}
Z&=&\int \,DU\,\exp\left[\sum_p\frac{\beta}{2N}\left( \mathrm{Tr}\, U+\mathrm{Tr}\,U^{\dagger}-2N\right)  \right], \\ 
\beta&=&\frac{2N}{g^2}.\nonumber
\end{eqnarray}
An expansion in the lattice coupling $\beta$ by group characters $\chi_r(U)$ and a cluster expansion yields the 
free energy density {\cite{Montvay:1994cy}} 
\begin{equation}
\tilde{f}\equiv-\frac{1}{\Omega}\ln Z=-6\ln\,c_0(\beta)-\frac{1}{\Omega}
\sum_{C=(X_i^{n_i})}\,a(C)\prod_i\Phi(X_i)^{n_i}.
\label{free}
\end{equation}
where $\Omega=V\cdot N_t$ is the lattice volume and $c_0$ is the expansion coefficient of the trivial representation, 
which has been factored out. The combinatorial factor $a(C)$ is introduced via a moment-cumulant-formalism, 
and equals $1$ for clusters $C$ which consist of only one so-called polymer $X_i$. 
The quantity in equation (\ref{free}) is customarily called a free energy, even at zero physical temperature, because
the path integral corresponds to a partition function if one 
formally identifies $\beta$ with
$1/T$. Here we are interested in a physical temperature $T=1/(aN_t)$, realized by compactifying the
temporal extension of the lattice. 
The physical free energy is then obtained by subtracting the formal ($N_t=\infty$) free energy, 
which is analogous to a subtraction of the divergent vacuum energy in the continuum. 
Thus the physical free energy density reads
\begin{equation}
f(N_t,u)=\tilde{f}(N_t,u)-\tilde{f}(\infty,u).
\end{equation}
The contributing polymers $X_i$ have to be objects with a closed surface, since 
\begin{eqnarray}
\int dU \chi_r(U)=\delta_{r,0}.
\end{eqnarray}
This means the group integration projects out the trivial representation at each link. To calculate the group integrals one uses the integration formula
\begin{equation}
\int dU \chi_r(UV)\chi_r(WU^{-1})=\chi_r(VW).
\end{equation}
For a more detailed introduction to strong coupling calculations we refer to {\cite{Montvay:1994cy}}.

\begin{figure}[t]
\vspace*{1cm}
\hspace*{-3cm}
\includegraphics[viewport= 60 700 660 690]{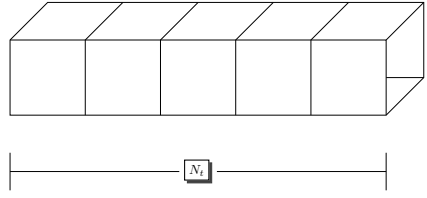}
\vspace*{2cm}
\caption{Graph contributing to the lowest order $f_1(N_t,u)$ of the expansion of the physical free energy density at finite temperature.}\label{fig_tube}
\end{figure}

\subsection{Results}

The graph contributing to the lowest order of the free energy density is a tube of length $N_t$ with a cross-section of one single plaquette. %(fig. \ref{fig_tube}). 
The contribution of these tubes together with inner plaquettes is
\begin{eqnarray}
\mbox{SU(2):}\qquad f_1(N_t,u)&=&-\frac{3}{N_t}u^{4N_t}a^{N_t},\\ 
\mbox{SU(3):}\qquad f_1(N_t,u)&=&-\frac{3}{N_t}u^{4N_t}\left[b^{N_t}+c^{N_t}\right],
\end{eqnarray}
where $u$, $v$ and $w$ are the expansion parameters of the 
lowest dimensional representations of the corresponding gauge groups,
\begin{eqnarray}
\mbox{SU(2):}\qquad u&=&\frac{\beta}{4}+{{\cal{O}}(\beta^{2})}\qquad v=\frac{\beta^2}{24}+{{\cal{O}}(\beta^{4})},\qquad \nonumber\\
\mbox{SU(3):}\qquad u&=&\frac{\beta}{18}+{{\cal{O}}(\beta^{2})}\qquad v=\frac{\beta^2}{432}+{{\cal{O}}(\beta^{4})}\qquad w=\frac{\beta^2}{288}+{{\cal{O}}(\beta^{4})},
\end{eqnarray}
and we have used the abbreviations
\begin{eqnarray}
a&=&1+3v-4u^2,\nonumber\\
b&=&1-3u-6v+8w,\nonumber\\
c&=&1+3u+6v+8w-18u^2.
\end{eqnarray}

Higher order contributions consist of such tubes with local decorations of additional plaquettes either in the fundamental or in higher representations. For the interesting cases SU(2) and SU(3), these contributions up to the calculated orders are
\begin{eqnarray}
\mbox{SU(2):}\quad f(N_t,u)=&-&\frac{3}{N_t}u^{4N_t}a^{N_t}\left[ 1+12N_tu^4-\frac{1556}{81}N_tu^6+\left(86N_t^2+\frac{35828}{243}N_t \right)u^8 \right], \\
\nonumber\\
\mbox{SU(3):}\quad f(N_t,u)=&-&\frac{3}{N_t}u^{4N_t}c^{N_t}\Big[1+12N_tu^4+42N_tu^5-\frac{115343}{2048}N_tu^6-\frac{1095327}{2048}N_tu^7\Big]\nonumber\\
\nonumber\\
&-&\frac{3}{N_t}u^{4N_t}b^{N_t}\Big[1+12N_tu^4+30N_tu^5-\frac{17191}{256}N_tu^6+63N_tu^7\Big],
\end{eqnarray}
which are valid only for $N_t\geq5$. For smaller $N_t$ there are modifications of these formulae 
coming from polymers with cross-sections larger than one plaquette. The complete results for $N_t=2$ and $3$ 
in SU(2) are
\begin{eqnarray}
N_t=2:\quad f(2,u)=&-&\frac{3}{2}u^{8}\left[ 1-4u^2+\frac{110}{3}u^4-\frac{58472}{405}u^6+\frac{64897681}{65610}u^8 \right],\\
N_t=3:\quad f(3,u)=&-&u^{12}\left[ 1-6u^2+50u^4-\frac{37966}{135}u^6+\frac{843898}{405}u^8 \right],
\end{eqnarray}

\subsection{Free energy density as a glueball gas}

Recognizing the first orders of the corresponding glueball masses (see \cite{Munster:1981es} and \cite{Seo:1982jh}) for SU(2)
\begin{eqnarray}
m(A_1^{++})&=&-4\ln\,u+2u^2-\frac{98}{3}u^4-\frac{20984}{405}u^6-\frac{151496}{243}u^8,\\
m(E^{++})&=&-4\ln\,u+2u^2-\frac{26}{3}u^4+\frac{13036}{405}u^6-\frac{28052}{243}u^8,
\end{eqnarray}
and SU(3)
\begin{eqnarray}
m(A_1^{++})&=&-4\ln\,u-3u+9u^2-\frac{27}{2}u^3-7u^4-\frac{297}{2}u^5+\frac{858827}{10240}u^6+\frac{47641149}{71680}u^7,\\
m(E^{++})&=&-4\ln\,u-3u+9u^2-\frac{27}{2}u^3+17u^4-\frac{153}{2}u^5+\frac{1104587}{10240}u^6+\frac{29577789}{71680}u^7,\\
m(T_1^{+-})&=&-4\ln\,u+3u+\frac{9}{2}u^3-\frac{98}{4}u^4+\frac{33}{4}u^5-\frac{36771}{1280}u^6+\frac{117897}{448}u^7,
\end{eqnarray}
one can write
\begin{eqnarray}
SU(2):\qquad f(N_t,u)&=&-\frac{1}{N_t}\left[e^{-m(A_1^{++})N_t} + 2e^{-m(E^{++})N_t} + {\cal{O}}(u^4)\right], \\
SU(3):\qquad f(N_t,u)&=&-\frac{1}{N_t}\left[e^{-m(A_1^{++})N_t} + 2e^{-m(E^{++})N_t} + 3e^{-m(T_1^{+-})N_t}+ {\cal{O}}(u^4)\right],
\end{eqnarray}
corresponding to a gas of non-interacting glueballs in a hadron-resonance-gas model \cite{Karsch:2003zq}, where
\begin{eqnarray}
f\simeq-T\sum_i e^{-\frac{E_i}{T}}.
\end{eqnarray}
This is a rather remarkable result. 
It allows to see from first principles that the pressure $p=-f$ is exponentially small in the confined phase, 
and it explains the success of the hadron-resonance-gas model in reproducing the confined phase equation of state. 
Since the partition function is not directly measurable in Monte-Carlo simulations,
the pressure is usually obtained by the integral method \cite{Boyd:1996bx}, where
the expectation values of derivatives are computed and then integrated numerically,
\begin{eqnarray*}
 \frac{p}{T^4}\,\bigg\vert_{\beta_0}^{\beta}=N_t^4\int_{\beta_0}^{\beta}d\beta' \left[ S_0-S_T\right], 
\end{eqnarray*}
with $S_0=6P_0$ and $S_T=3(P_t+P_s)$, where $P_0$ denotes the plaquette expectation value on symmetric lattices and $P_{t,s}$ are those of space-time and space-space plaquettes with $N_t<N_s$.
The lower integration limit is usually set to zero by hand, arguing with an exponentially small pressure in the 
low temperature regime. Our results now justify this assumption from first principles.

\subsection{Phase transition}

Physical phase transitions limit the radius of convergence on the real $\beta$-axis, 
signalled by a singularity in the full free energy. We model the full function from the series coefficients
by Pad\'e approximants $[L,M]$ with
\begin{eqnarray*}
[L,M](u)\equiv \frac{1+a_1u+\dots +a_Lu^L}{b_0+b_1u+\dots+b_Mu^M},
\end{eqnarray*}
and search for the zeroes of the denominator. 
The resulting $L+M=2,3,4$ Pad\'e tables for $N_t=2,3$ with the nearest 
real singularities are shown in table (\ref{pade}). 

\begin{table}[b]
\begin{center}
\begin{minipage}{6.1cm}
{\bf{SU(2):  $N_t=2$}}\\[1ex]
\begin{tabular}{|c|c|c|c|}\hline
$[L,M]$ &  $u_c$     &  $\beta_c$ & $|u_c-u_0|$  \\ \hline\hline
$[1,2]$ &  $0.4033 $ &  1.8227    & 0.0899       \\
$[0,3]$ &  $0.4675 $ &  2.2201    &              \\
$[2,2]$ &  $0.5201 $ &  2.5981    &              \\
$[1,3]$ &  $0.4684 $ &  2.2262    &              \\ 
$[0,4]$ &  $0.4684 $ &  2.2261    &              \\ 
\hline
\end{tabular}
\end{minipage}
\begin{minipage}{6.1cm}
{\bf{SU(2):  $N_t=3$}}\\[1ex]
\begin{tabular}{|c|c|c|c|}\hline
$[L,M]$ &  $u_c$     &  $\beta_c$ & $|u_c-u_0|$  \\ \hline\hline
$[1,2]$ &  $0.3467 $ &  1.5133    & 0.0219       \\
$[0,3]$ &  $0.5009 $ &  2.4538    &              \\
$[2,2]$ &  $0.4623 $ &  2.1853    & 0.2388       \\
$[1,3]$ &  $0.4347 $ &  2.0098    & 0.1373       \\ 
$[0,4]$ &  $0.4617 $ &  2.1820    &              \\ 
\hline
\end{tabular}
\end{minipage}
\end{center}
\caption[]{Zeroes of the denominator ($u_c$) and the numerator ($u_0$) of the 
$[L,M]$ Pad\'e approximants and the corresponding value of $\beta_c$.}
\label{pade}
\end{table}

\begin{figure}[t]
\vspace*{-1cm}
\begin{minipage}{7cm}
{\includegraphics[width=7cm]{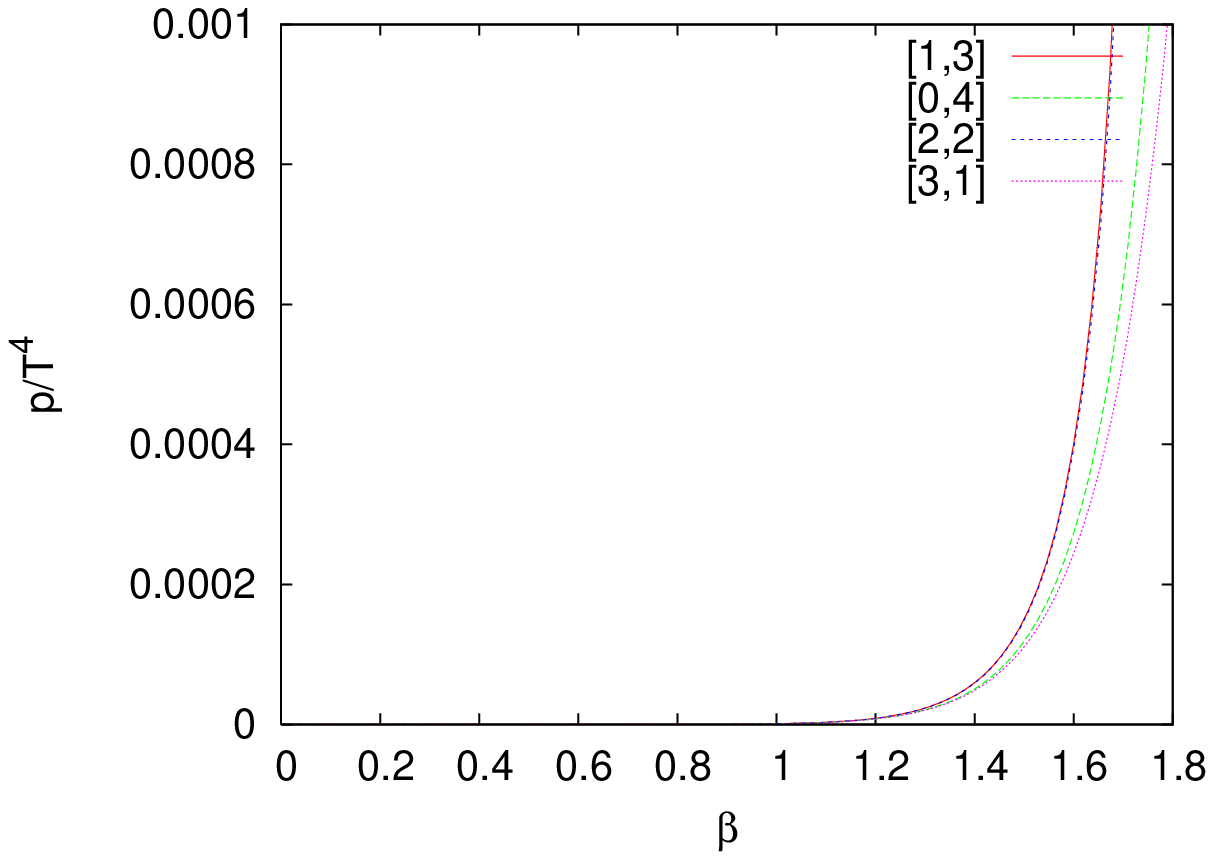}}
\end{minipage}
\begin{minipage}{5cm}
\includegraphics[width=5cm]{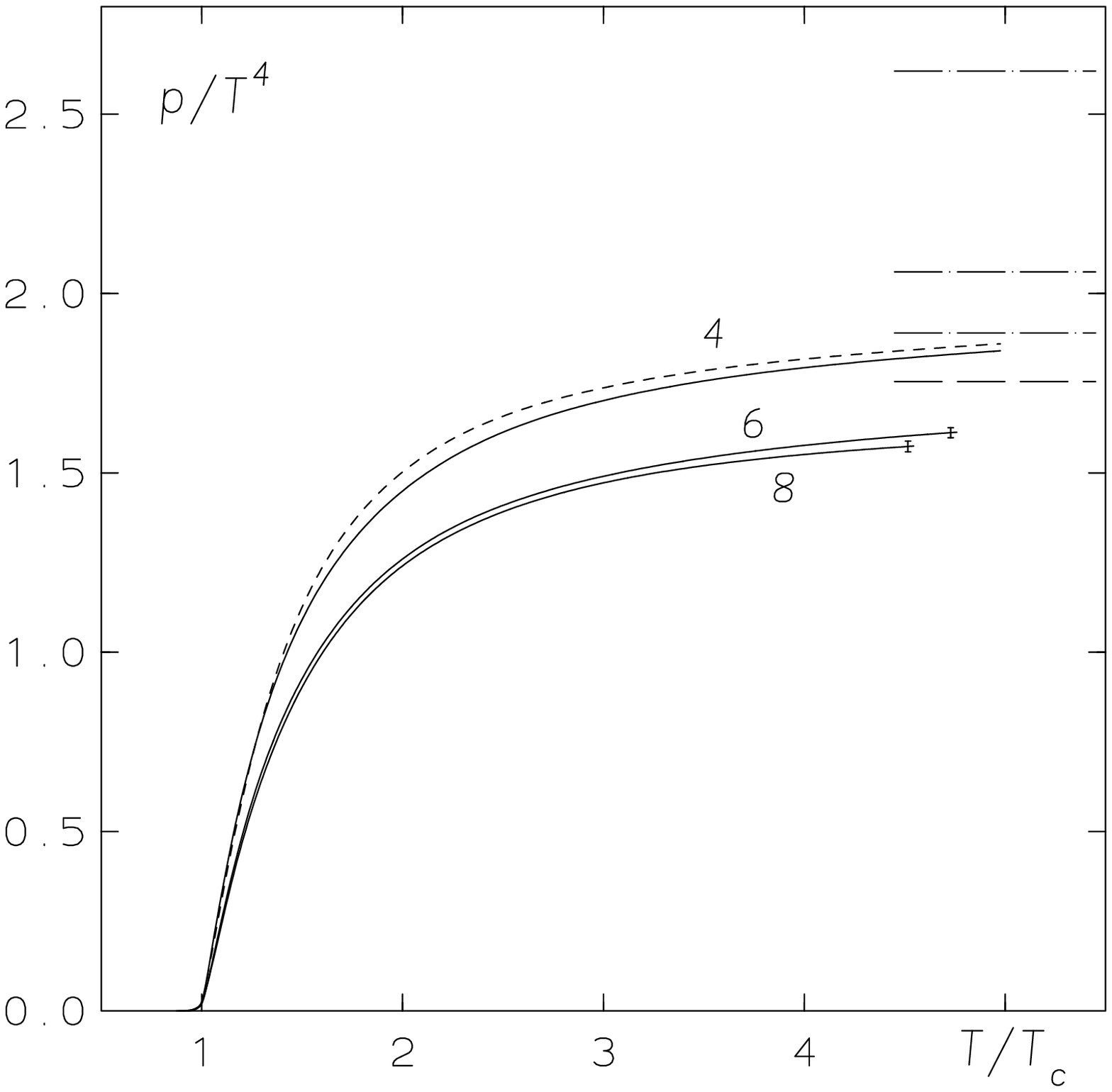}
\end{minipage}
\caption{Left: Plot of the pressure density $p$ vs. $\beta$ ($L+M=4$ Pad\'es for SU(2) and $N_t=3$).
The plot range corresponds to the confined phase up to the critical coupling.
Right: Plot of $p$ vs. $T/T_c$ for SU(3) from Monte Carlo data \protect\cite{Boyd:1996bx}.}
\label{fig_pressure}
\end{figure}

Zeroes $u_0$ of the Pad\'e approximant which are very close to a singularity often indicate
that the singularity is superfluous and disappears as the full funtion is approached. 
Hence, removing the singularities with a nearby zero, we obtain estimates for the critical couplings, which 
are not far from the Monte Carlo results $\beta_c=1.8800(30)$ for $N_t=2$ and $\beta_c=2.1768(30)$
for $N_t=3$ \cite{Fingberg:1992ri}. 

The $L+M=4$ Pad\'e's for $N_t=3$, SU(2), are shown in figure (\ref{fig_pressure}).
 The spread in the curves
gives an estimate of the systematic error of the approximants at that order. 
The exponential suppression in the confined phase as well as the onset of the pressure
upon approaching $T_c$
is reproduced by the strong coupling series.

\section{Screening masses}

\subsection{Zero temperature}

Screening masses are defined by the exponential decay of the spatial correlation of suitable operators. 
We used plaquette operators in our calculations. Temporarily assigning separate gauge couplings to 
all plaquettes, the correlator can be defined as \cite{Munster:1981es} 
\begin{equation}
C(z)=\langle\mathrm{Tr}\,U_{p_1}(0)\,\,\mathrm{Tr}\,U_{p_2}(z)
\rangle=N^2\frac{\partial^2}{\partial\beta_1\partial\beta_2}\ln\,
Z(\beta,\beta_1\beta_2)\bigg\vert_{\beta_{1,2}=\beta}.
\end{equation}
At zero temperature the exponential decay is the same as for correlations in the time direction, and thus
determined by the gluball masses, the lowest of which may be extracted as
\begin{equation}
m=-\lim_{z\rightarrow\infty}\frac{1}{z}\ln\,C(z).\\
\end{equation}
The leading order graphs for the strong coupling series are shown in figure (\ref{mass}). 
This leads to the lowest order contribution:
\begin{equation}
C(z)=A\,u^{4z}=A\mathrm{e}^{-m_sz}.
\end{equation}
Thus the leading order for the screening mass is given by
\begin{equation}
m_s=-4\ln\,u(\beta).
\end{equation}

\subsection{Finite temperature}

\begin{figure}[t]
\vspace*{1cm}
\hspace*{-1cm}
\includegraphics[viewport= 60 700 660 690]{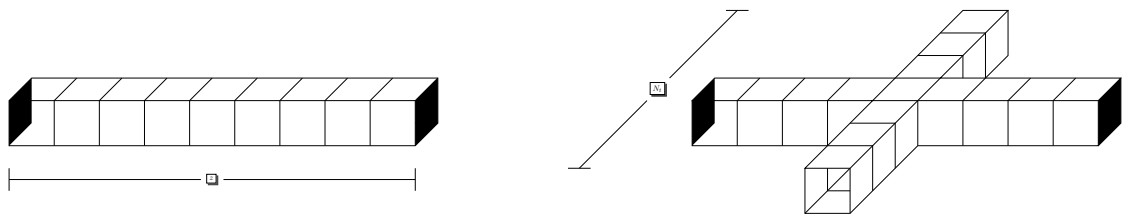}
\vspace*{2cm}
\caption{Graphs contributing to the lowest order of the expansion of the screening mass at vanishing and finite temperature. The correlation-plaquettes are painted black.}\label{fig_scrmass}
\label{mass}
\end{figure}

The graph contributing to the lowest order of the difference between the 
screening masses at zero and finite temperature is shown in 
figure (\ref{fig_scrmass}).
To lowest order the mass difference is
\begin{eqnarray}
\Delta m_s(N_t)&=&m_s(N_t)-m_s(\infty)\\
\\
&=&-\frac{2}{3}N_tu^{4N_t-6}
\end{eqnarray}
Thus one can see that the finite temperature effect on the screening 
mass is very small below $T_c$, 
as is also observed in Monte Carlo simulations (for references, see \cite{Laermann:2003cv}).

\section{Conclusions}

We performed explorative studies of strong coupling expansions at finite temperature. Our series for the 
free energy density is to the lowest orders consistent with a free glueball gas. 
This result justifies the neglect of the lower
integration constant in numerical calculations of the equation of state 
by the integral method from first principles. 
Moreover, it gives an explanation for the success of the hadron-resonance-gas model in reproducing lattice data in 
the confined phase.
By extrapolating the power series via Pad\'e approximants and looking for the zeroes of the denominator, 
it is possible to get estimates for the critical value $\beta_c$ of the deconfining phase transition, although
higher order terms seem necessary in order to obtain some accuracy here.
Finally, glueball screening masses show a weak temperature dependence in the confined phase, consistent with what
is found in numerical simulations.

\end{document}